\documentclass[prl,twocolumn]{revtex4}
\usepackage{type1cm}
\usepackage[dvipdfm]{graphicx}

\usepackage{amsmath,amssymb,bm,amsthm}

\def\beq{\begin{equation}}
\def\eeq{\end{equation}}
\def\nbeq{\begin{equation*}}
\def\neeq{\end{equation*}}

\def\<{\langle}
\def\>{\rangle}

\renewcommand{\d}{\partial}

\begin{document}
\title{Nonadditivity in Quasiequilibrium States of Spin Systems with Lattice Distortion}

\author{Takashi Mori}
\email{
mori@spin.phys.s.u-tokyo.ac.jp}
\affiliation{
Department of Physics, Graduate School of Science,
University of Tokyo, Bunkyo-ku, Tokyo 113-0033, Japan
}

\begin{abstract}
It is argued that a certain kind of short-range interacting system exhibits nonadditivity when several time scales are well separated.
Under the condition of separated time scales, the system is described by the elastic spin model.
We find that it is extensive but nonadditive, 
which is directly confirmed by the work measurement and also indicated by ensemble inequivalence.
Further, we estimate the effective Hamiltonian for the spin variables, and it is clarified that the effective interaction is long ranged.
Remarkably, the so-called Kac prescription, which is usually regarded as a mathematical operation to make the system extensive, 
naturally holds.
\end{abstract}
\maketitle

Let us consider a system consisting of two macroscopic subsystems $A$ and $B$.
In a usual macroscopic system, the total amount of energy is given by the sum of internal energies of the two subsystems
because the interaction energy between $A$ and $B$ is negligible compared to the bulk energy.
This property is called additivity (the precise definition will be given later).
Additivity is regarded as a fundamental property of macroscopic systems~\cite{Callen_text}.
It ensures concavity or convexity of the thermodynamic function.
In statistical mechanics, it leads to the ensemble equivalence; i.e.,
several statistical ensembles yield identical thermodynamic quantities~\cite{Ruelle_text}.
However, not all the macroscopic systems possess additivity.
Long-range interacting systems are representative of nonadditive and physically relevant systems~\cite{Campa_review2009,Les_Houches2008}.
Because of the lack of additivity, long-range interacting systems can exhibit unfamiliar and peculiar macroscopic properties,
e.g., negative specific heat~\cite{Lynden-Bell-Wood1968,Thirring1970}, ensemble inequivalence~\cite{Thirring1970}, 
macroscopic inhomogeneity~\cite{Mori2011_instability},
and no thermalization in an isolated system~\cite{Antoni-Ruffo1995}.

Apparently, a short-range interacting system is unlikely to be nonadditive 
since the interaction energy will be very small compared to the bulk energy.
In this Letter, however, it is pointed out that in a {\it quasi}equilibrium state, a certain kind of short-range interacting system can exhibit nonadditivity.
Interestingly, in spite of its nonadditivity, the system is extensive;
the energy is proportional to the system size if we make the system large uniformly.

Now we explain the elastic spin model studied in this Letter.
The model itself has already been known in studies on spin-crossover materials~\cite{Nishino2007}.
See Ref.~\cite{Gutlich_review1994} and references therein for a detailed account of spin-crossover materials.
$N$ molecules are aligned on the two-dimensional triangular lattice of side $L$, where $N=L^2$.\footnote
{The conclusion of this work does not change in three dimensions.
In the one-dimensional chain, however, it is shown that the elastic spin model does not exhibit nonadditivity;
see K. Boukheddaden, S. Miyashita, and M. Nishino, Phys. Rev. B 75, 094112 (2007).}
Each molecule is composed of a metal ion and surrounding ligands.
Each molecule $i$ has two different stable internal states, that is, the high-spin (HS) state $\sigma_i=+1$ and the low-spin (LS) state $\sigma_i=-1$.
Electron configurations in a metal ion are different between the HS and the LS states
and the HS state has a higher spin than the LS state.
For example, in $\rm Fe^{II}$, the HS state has $S=2$ and the LS state has $S=0$.
As a result, the HS state with $S=2$ has degeneracies of $S^z=2,1,0,-1,-2$.
In general, degeneracies of the HS state are denoted by a parameter $g$ and thus
$\sigma_i\in\{\underbrace{+1,+1,+1,\dots,+1}_{g},-1\}.$
Although $\sigma_i$ is not a genuine spin, we call it a spin variable and $M=\sum_i^N\sigma_i$ the magnetization.
The magnetization density is denoted by $m\equiv M/N$.

\begin{figure}[t]
\begin{center}
\begin{tabular}{cc}
(a)&(b)\\
\includegraphics[clip,width=4cm]{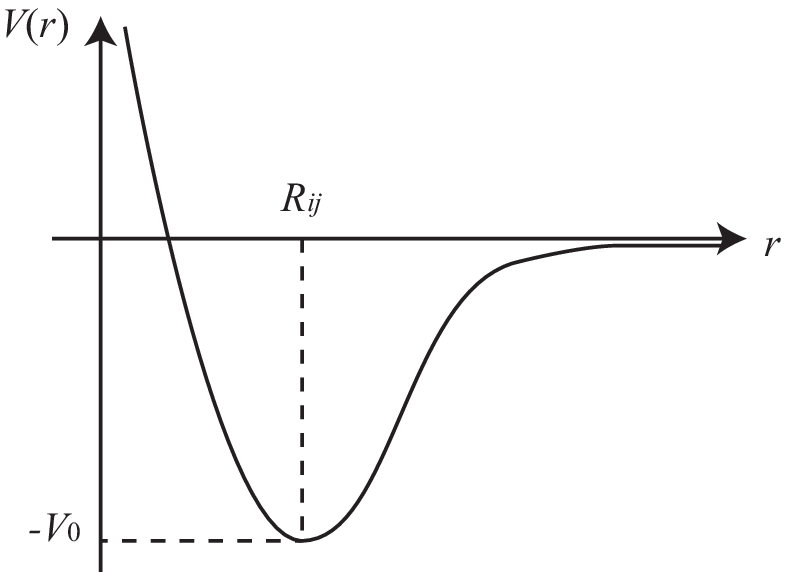}&
\includegraphics[clip,width=4cm]{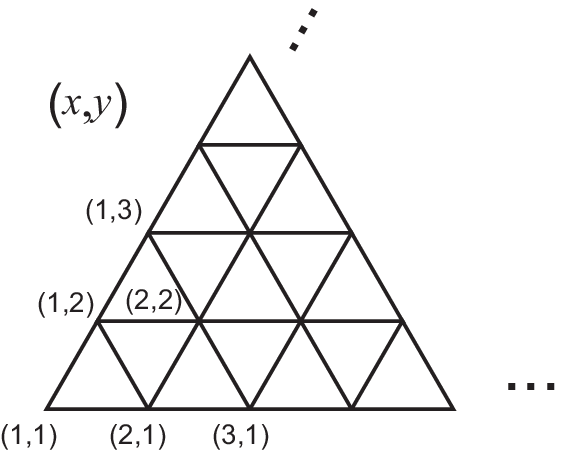}
\end{tabular}
\caption{(a) Typical shape of the interaction potential.
(b) Triangular lattice and the label of the molecules.}
\label{fig:setting}
\end{center}
\end{figure}

The intermolecular interaction is given by some short-range potential $V_{ij}(r)$ such as Fig.~\ref{fig:setting}(a), which decays faster than $1/r^2$ in a long distance.
Because of the short-range nature of the interaction, it is sufficient to consider only the nearest-neighbor interactions.
The important point is that the equilibrium distance of $V_{ij}(r)$ is given by $R_{ij}\equiv R(\sigma_i)+R(\sigma_j)$,
which depends on the molecular internal states.
The values of $R(\pm1)$ represent the molecular radius at state $\sigma_i=\pm1$, respectively.
This size difference is actually observed in experiments~\cite{Wiehl1990} and
plays an important role for spin-crossover transitions~\cite{Miyashita2008}.

When the potential depth $V_0$ in Fig.~\ref{fig:setting}(a) is much larger than the thermal energy $k_{\rm B}T$, 
where $k_{\rm B}$ is the Boltzmann constant and $T$ is the temperature, $V_{ij}(r)$ is approximated as a quadratic form:
\beq
V_{ij}(r)\simeq \frac{k}{2}(r-R_{ij})^2=\frac{k}{2}\left\{r-[R(\sigma_i)+R(\sigma_j)]\right\}^2.
\label{eq:approx}
\eeq
The condition of the applicability of this approximation is discussed later.

In this way, the Hamiltonian of the elastic spin model is given by
\beq
H=\sum_{i=1}^N\frac{\bm{p}_i^2}{2}
+\frac{k}{2}\sum_{\< i,j\>}\left(|\bm{q}_i-\bm{q}_j|-R_{ij}\right)^2
+D\sum_{i=1}^N\sigma_i.
\label{eq:elastic_H}
\eeq
Here $\bm{q}_i$ and $\bm{p}_i$ are the coordinate and the momentum of the $i$th molecule.
The symbol $\<i,j\>$ denotes all the nearest-neighbor pairs.
The last term represents the effect of the ligand field.
When $D=0$, there are two ground states: i.e., $\sigma_i=1$ $\forall i$ and $\sigma_i=-1$ $\forall i$.
Throughout this Letter, according to the previous work~\cite{Miyashita2008}, we fix the parameters as
$R(-1)=1$, $R(1)=1.1$, and $k=40.$

For convenience, we also label the molecules by the two-dimensional vectors $\bm{r}_i=(x_i,y_i)$
with $x_i,y_i\in\{1,2,\dots,L\}$.
The molecule at the $x$th column and $y$th row is labeled by $(x,y)$: see Fig.~\ref{fig:setting}(b).
We distinguish $\bm{q}_i$ from $\bm{r}_i$; the former is a dynamical variable denoting the position of the $i$th molecule,
but the latter is the label of the site denoting its positional relation on the lattice.

Here we should mention the condition under which we can consider the equilibrium state of Eq.~(\ref{eq:elastic_H});
the observation time $t_{\rm obs}$ should satisfy
\beq
\tau_{\rm eq}< t_{\rm obs}\ll\tau_V\propto\exp[V_0/k_{\rm B}T],
\label{eq:condition}
\eeq
where $\tau_{\rm eq}$ is the equilibration time of this model and $\tau_V$ is the activation time necessary to get over the potential depth $V_0$.
If this condition were violated, fracture of the solid would occur and the approximation (\ref{eq:approx}) is not necessarily valid;
it becomes important to consider molecules which are far away from the nearest neighbors.
When $V_0\gg k_{\rm B}T$, such slow things have not happened in the time scale $t_{\rm obs}$ given above,
and hence the model (\ref{eq:elastic_H}) is appropriate.
In other words, {\it an equilibrium state of the elastic spin model is regarded as a quasiequilibrium state of the original short-range interacting system}.

\begin{figure}[t]
\begin{center}
\includegraphics[clip,width=6cm]{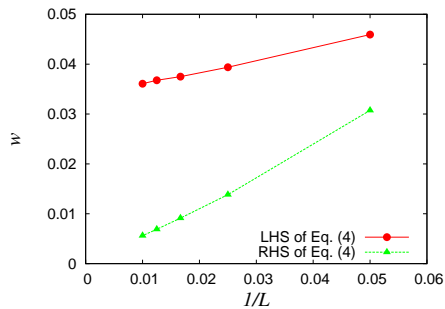}
\caption{(color online). lhs (red circles) and rhs (green triangles) of Eq.~(\ref{eq:additivity}) vs $1/L$.
In the former, we put $m_{A}=-1$ and $m_{B}=+1$.
In the latter, there is no restriction on $m_{A}$ and $m_{B}$.}
\label{fig:ext_add}
\end{center}
\end{figure} 

It is noted that the Hamiltonian given by Eq.~(\ref{eq:elastic_H}) is nonlocal and possesses long-range nature.
The essential point is that (i) the variables $\{\bm{q}_i\}$ are not independent 
since two nearest-neighbor molecules cannot be far apart within the time scale~(\ref{eq:condition})
and (ii) equilibrium positions of molecules depend on the spin variables $\{\sigma_i\}$ nonlocally,
which is due to the size difference between the HS and the LS states.

In order to investigate extensivity and additivity,
let us consider the system in contact with a thermal bath at the temperature $T$.
We virtually divide the system into two subsystems $A$ and $B$;
the subsystem $A$ is composed of the molecules with $x_i\leq L/2$ and the subsystem $B$ with $x_i>L/2$.
If the molecule $i$ belongs to the subsystem $A$ ($B$), we write $i\in A (B)$.
We divide the Hamiltonian into three parts, $H_{AB}(\lambda)=H_{A}+H_{B}+\lambda H_{I}$.
Here $H_{X}\equiv\sum_{i\in X}(\bm{p}_i^2/2+D\sigma_i)+(k/2)\sum_{\< i,j\> \in X}(|\bm{q}_i-\bm{q}_j|-R_{ij})^2$,
where $X$ is $A$ or $B$.
The parameter $\lambda$ controls the strength of the interaction between subsystems and
we choose it so that $H_{AB}(\lambda=1)$ is equal to Eq.~(\ref{eq:elastic_H}).

The concept of additivity is related to the probability that the magnetization of $A$ is equal to $M_{A}$ and that of $B$ is equal to $M_{B}$.
In this Letter, the system is said to be additive if $P_{AB}(T,M_{A},M_{B})\simeq P_{A}(T,M_{A})P_{B}(T,M_{B})$.
Here $P_{AB}(T,M_{A},M_{B})$ denotes the probability 
that the magnetization of $A$ at temperature $T$ is equal to $M_{A}$ and that of $B$ is equal to $M_{B}$ 
in a composite system $H_{AB}(1)$.
On the other hand, $P_{X}(T,M_{X})$ ($X$ is $A$ or $B$) denotes the probability 
that the magnetization of $X$ at temperature $T$ is equal to $M_{X}$ {\it in a decoupled system} $H_{X}$.

According to statistical mechanics, such probabilities are related to the thermodynamic functions with some restriction~\cite{Landau_stat}.
That is, $P_{AB}(T,M_{A},M_{B})=\exp\{[F_{AB}(T)-F_{AB}(T,M_{A},M_{B})]/k_{\rm B}T\}$,
where $F_{AB}(T)$ is the free energy of the composite system $H_{AB}(1)$ at temperature $T$ 
and $F_{AB}(T,M_{A},M_{B})$ is that under the constraint 
that the magnetizations of $A$ and $B$ are fixed to be $M_{A}$ and $M_{B}$, respectively.
Similarly, $P_{X}(T,M_{X})=\exp\{[F_{X}(T)-F_{X}(T,M_{X})]/k_{\rm B}T\}$, 
where $F_{X}(T)$ is the free energy associated with the Hamiltonian $H_{X}$ and $F_{X}(T,M_{X})$ is that  
under the constraint that its magnetization is fixed to be $M_{X}$.

Thus the condition of additivity is rewritten as 
\begin{align}
F_{AB}(T,M_{A},M_{B})-F_{A}(T,M_{A})-F_{B}(T,M_{B})
\nonumber \\
\simeq F_{AB}(T)-F_{A}(T)-F_{B}(T)
\label{eq:additivity}
\end{align}
for any values of $M_{A}$ and $M_{B}$.
In particular, when the right-hand side (rhs) of Eq.~(\ref{eq:additivity}) is zero, the system is said to be extensive.
According to thermodynamics, the left-hand-side (lhs) of Eq.~(\ref{eq:additivity}) 
is equal to the amount of work done by the system in a quasistatic (reversible) isothermal process of changing $\lambda$ from 1 to 0.
During the process, $M_{A}$ and $M_{B}$ should be individually conserved.

In Fig.~\ref{fig:ext_add}, we calculated the work per molecule, $w=-\int\<\d H(\lambda)/\d\lambda\>\dot{\lambda}dt/L^d$, 
in such a quasistatic isothermal process.
In the red circles, we put $m_{A}=M_{A}/L^d=-1$, $m_{B}=M_{B}/L^d=+1$, and $T=0.1$. 
The value of $w$ corresponds to the lhs of Eq.~(\ref{eq:additivity}).
It is observed that $w$ does not vanish as $L$ increases.
On the other hand, the green triangles are measured values of $w$ without restriction on the magnetizations, 
which correspond to the rhs of Eq.~(\ref{eq:additivity}).
In this case $w$ tends to zero as $L$ increases.
Therefore, it is concluded that the system is extensive but nonadditive.

\begin{figure*}[t]
\begin{center}
\begin{tabular}{ccc}
(a)&(b)&(c)\\
\includegraphics[clip,width=5cm]{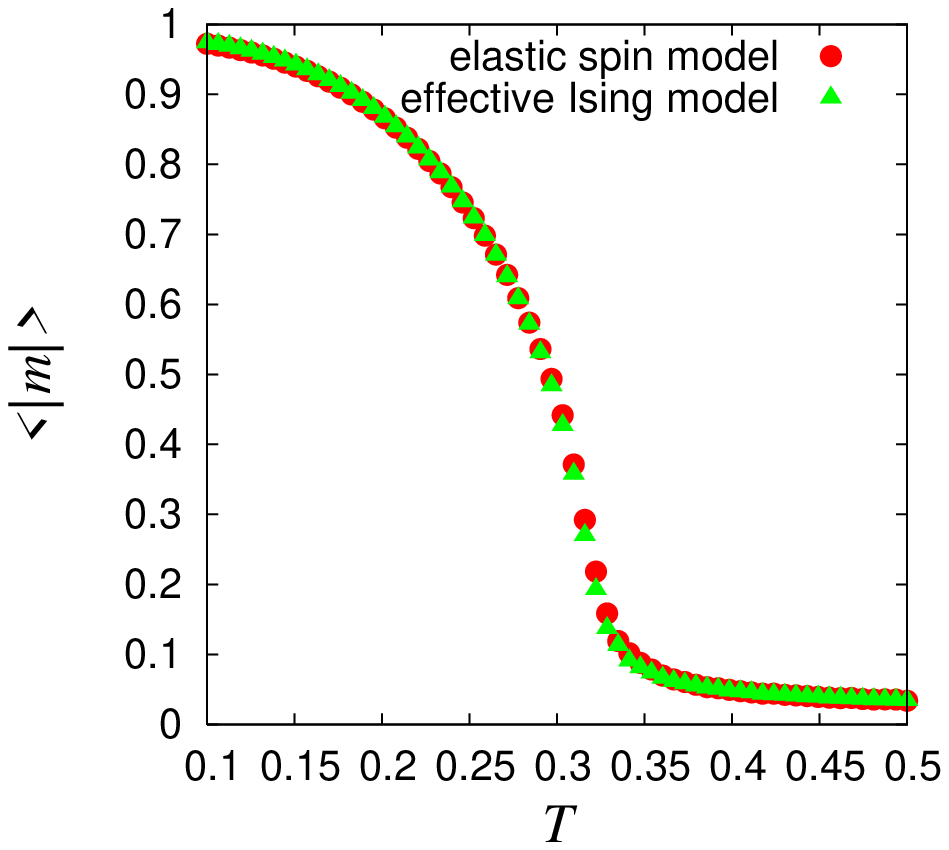}&
\includegraphics[clip,width=5cm]{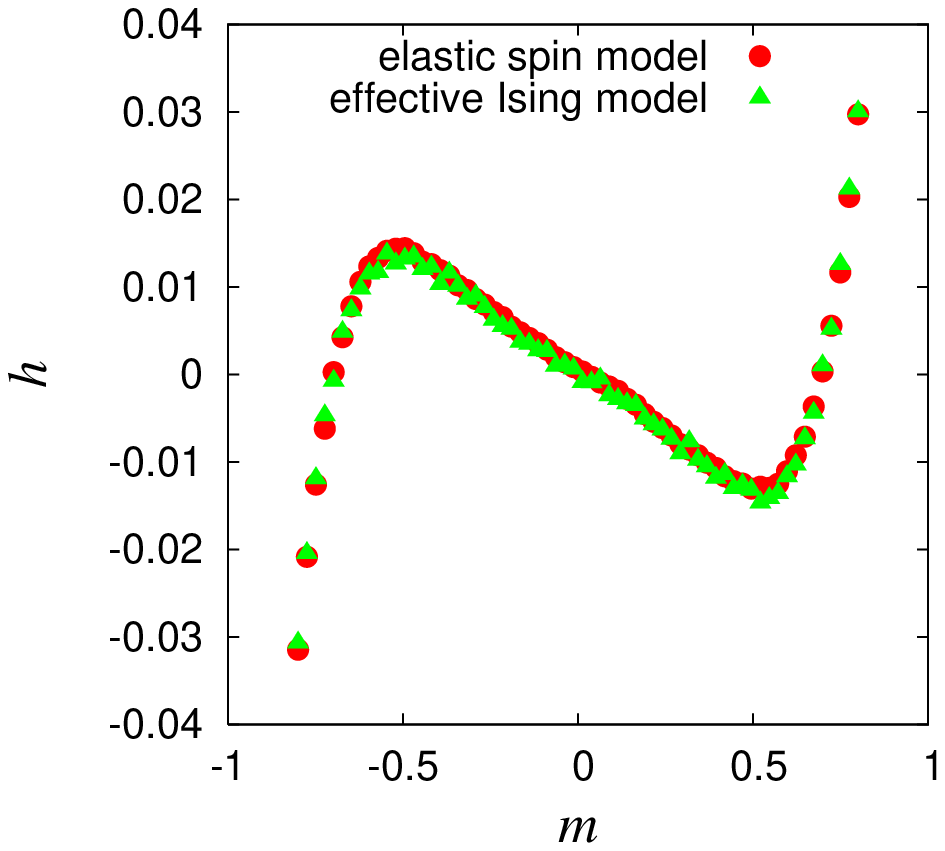}&
\includegraphics[clip,width=5cm]{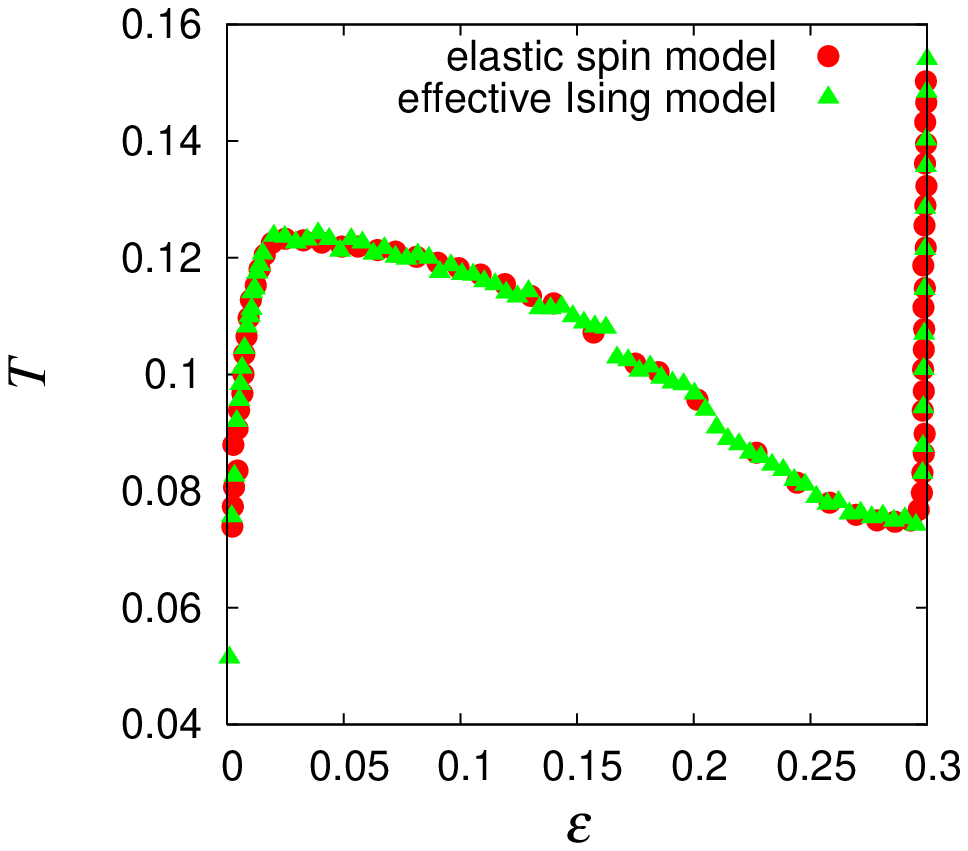}
\end{tabular}
\caption{(color online). Equilibrium quantities.
The red circles and the green triangles represent the results for the elastic spin model and for the effective Ising model, respectively.
(a) Absolute magnetization densities $\< |m|\>$ are plotted against the temperature in the canonical ensemble.
(b) Magnetic fields are plotted against magnetization densities at $T=0.26$ in the restricted canonical ensemble.
(c) Temperatures are plotted against energy densities in the microcanonical ensemble.
In order to compare the two models, the contribution of the lattice vibration, $2T$, is subtracted from the energy densities for the elastic spin model.}
\label{fig:elastic}
\end{center}
\end{figure*} 

Because of nonadditivity, it is expected that the elastic spin model exhibits the peculiar properties observed in long-range interacting systems.
We performed Monte Carlo simulations
for (a) the canonical ensemble, (b) the restricted canonical ensemble (the canonical ensemble with restriction on the value of the magnetization),
and (c) the microcanonical ensemble.
Numerical results are shown by the dark gray (red) circles in Fig.~\ref{fig:elastic}.
In the microcanonical ensemble, we subtracted $2T$, which represents the contribution of the lattice vibration, from the total energy density.
The parameters are set to be $D=0$ and $g=1$ for (a) and (b), and $D=0.15$, $g=20$ for (c) \footnote
{In the present model, the negative specific heat does not appear at $D=0$ and $g=1$.
In order to demonstrate nontrivial behavior of this model, we chose parameters as $D=0.15$ and $g=20$ in (c).}.
The system size is $L=40$.

In the canonical ensemble, spin variables are changed according to the usual Metropolis algorithm
and the positions of molecules move according to the Hamilton dynamics (by the leapfrog algorithm).
We can see that the system undergoes a second order phase transition at $T_{\rm c}\simeq 0.35$ in Fig.~\ref{fig:elastic}(a).
The critical behavior belongs to the mean-field universality class~\cite{Miyashita2008} (see also Ref.~\cite{Nakada2012}).

The algorithm used in the restricted canonical ensemble is the same as in the canonical ensemble 
except that we prepare an excess degree of freedom referred to as the ``demon''.
The demon keeps the magnetization $m_{d}=\pm 1$.
Only if $m_{d}\sigma_i\leq 0$, the spin flip is accepted according to the Metropolis transition probability.
After the flip, we change $\sigma_i$ and $m_{d}$ to $-\sigma_i$ and $-m_{d}$, respectively.
The advantage of this method is that we can easily measure the magnetic field $h=\d F(T,M)/\d M$ from the average value of $m_{d}$ by
$h=\frac{k_{\rm B}T}{2}\ln\frac{1+\<m_d\>}{1-\<m_d\>}$.
In Fig.~\ref{fig:elastic}(b), we can clearly see the region where the susceptibility $\chi=\d m/\d h$ is negative.
The susceptibility is always positive in the canonical ensemble 
and this discrepancy shows that the canonical ensemble is inequivalent to the restricted canonical ensemble.

In the microcanonical ensemble,
we used the leapfrog algorithm for the time evolution of $\{\bm{q}_i,\bm{p}_i\}$
and the Creutz algorithm~\cite{Creutz1983} for the spin flip, in which an excess degree of freedom also called the demon, which has a positive energy, is prepared
and the total energy of the system and the demon is conserved.
The temperature of the system can be measured simply by $T=\<E_{d}\>$.
Figure~\ref{fig:elastic}(c) clearly demonstrates that there is a region of the negative specific heat, which also shows the ensemble inequivalence.

Although there is no direct interaction between $\sigma_i$ and $\sigma_j$ in long distance,
it will be a plausible consideration that an {\it effective interaction} arises between spin variables via lattice distortion.
In general, the effective Hamiltonian for $\{\sigma_i\}$ obtained by eliminating $\{\bm{q}_i,\bm{p}_i\}$ 
contains many-body interactions and will be very complicated.
Here we assume that the effective Hamiltonian is written as $H_{\rm eff}=-(1/2)\sum_{ij}\hat{J}_{ij}\sigma_i\sigma_j$
and we try to estimate $\hat{J}_{ij}$ from the numerical data of correlation functions $\hat{C}_{ij}\equiv\<\sigma_i\sigma_j\>$.
We put $g=1$ and $D=0$ and consider the canonical ensemble in the high-temperature phase ($T=0.5$).
In this case, we can show that $\hat{J}_{ij}$ is obtained from $\hat{C}_{ij}$ by
\beq
\hat{C}(L)=[\hat{I}-\hat{J}(L)/(k_{\rm B}T)]^{-1},
\label{eq:C_J}
\eeq
where the identity matrix is denoted by $\hat{I}$.
The relation~(\ref{eq:C_J}) is approximate one in general, but it becomes exact if the effective interaction is long ranged: i.e.,
\beq
\hat{J}_{ij}(L)=L^{-d}\phi(L^{-1}\bm{r}_{ij}),
\label{eq:J_scaling}
\eeq
where $\bm{r}_{ij}=(x_{ij},y_{ij})\equiv\bm{r}_i-\bm{r}_j$ and $d$ is the spatial dimension (here $d=2$).
It is assumed that $\int_{|\bm{x}|<\delta}\phi(\bm{x})d^dx<+\infty$ for an arbitrary fixed $\delta>0$.
It includes the power-law interaction $\phi(\bm{x})\sim 1/x^{\alpha}$ with $\alpha<d$.
The scaling $L^{-1}\bm{r}_{ij}$ means that the interaction range is comparable with the system size $L$.
The scaling $L^{-d}$ in Eq.~(\ref{eq:J_scaling}) ensures that the energy is proportional to the system size,
which corresponds to the so-called Kac prescription~\cite{Campa_review2009,Les_Houches2008}.

\begin{figure}[tb]
\begin{center}
\begin{tabular}{cc}
(a)&(b)\\
\hspace{-3em}\includegraphics[clip,width=5cm]{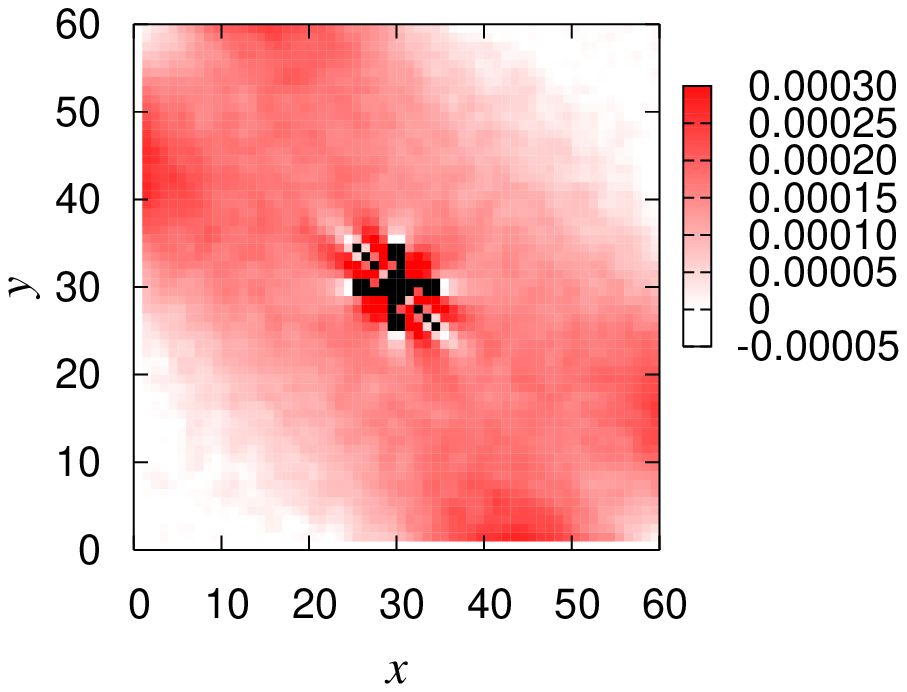}&
\hspace{-3em}\includegraphics[clip,width=5cm]{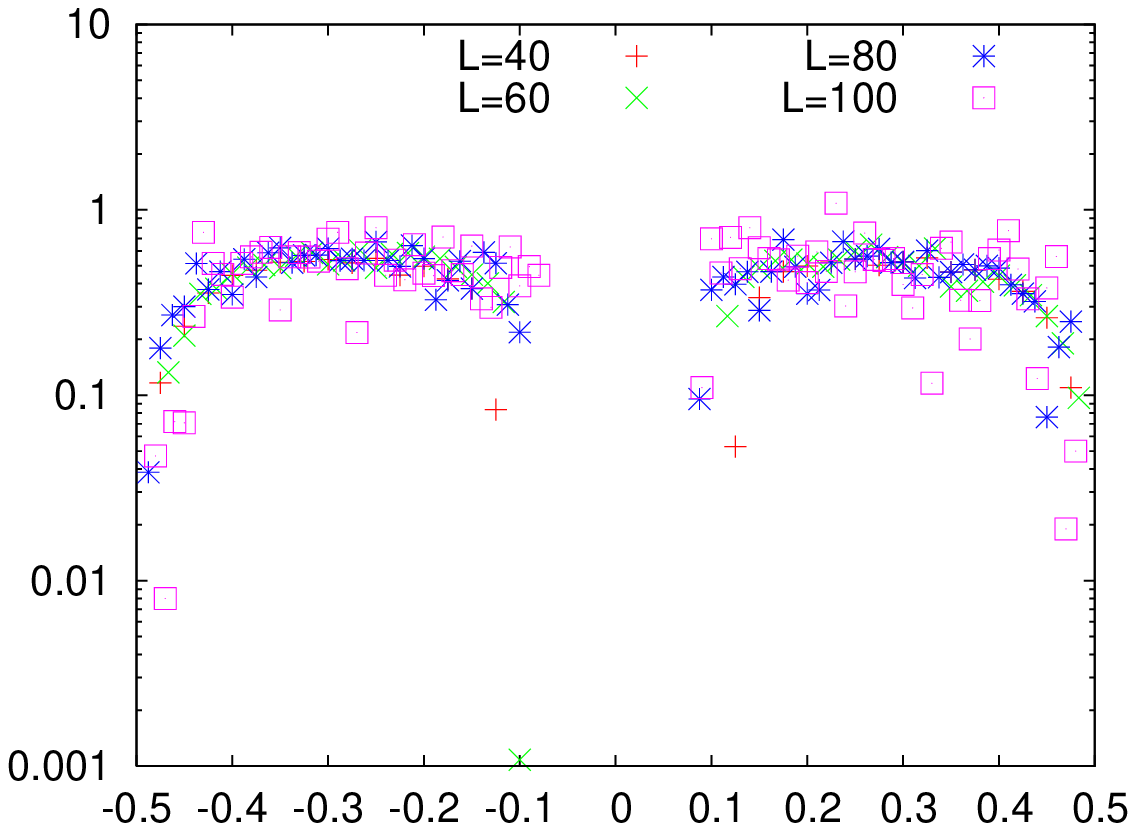}
\end{tabular}
\caption{(color online). (a) Global image of the interaction potential at $T=0.5$ and $L=60$.
The site $j$ is fixed at the center of the system ($x_j=y_j=30$).
In the white region, the interaction is almost zero; $-0.00005<\hat{J}_{ij}\leq 0$.
In the black region, the interaction is antiferromagnetic; $\hat{J}_{ij}\leq -0.00005$.
In the gray (red) region, the interaction is ferromagnetic and the depth of the color expresses the strength of the ferromagnetic interaction.
(b) Graphs of the estimated effective interactions along the diagonal direction, $y_{ij}=L-x_{ij}$ for $L=40,60,80,$ and 100.
The transverse axis is the scaled distance $x_{ij}/L$ and the longitudinal axis is the scaled interaction matrix $L^2\hat{J}_{ij}$.
Only the points with $\hat{J}_{ij}>0$ are plotted.}
\label{fig:interaction_L}
\end{center}
\end{figure}

In Fig.~\ref{fig:interaction_L}(a), the structure of the estimated interaction matrix $\hat{J}_{ij}$ is depicted for $L=60$.
In the figure, the central site $\bm{r}_j=(L/2,L/2)$ is chosen as the site $j$.
It is found that the interaction is highly anisotropic and a long-range ferromagnetic interaction emerges.
Importantly, its spatial average does not vanish;
the long-range ferromagnetic interaction is not screened.

The characteristic system size dependence of the estimated interaction is shown in Fig.~\ref{fig:interaction_L}(b),
where the graphs of $L^d\hat{J}_{ij}(L)$ along the diagonal direction, $y_{ij}=L-x_{ij}$, are depicted 
as a function of $L^{-1}x_{ij}$ for several values of $L$ only in the region of $\hat{J}_{ij}>0$.
We can see that these graphs are collapsed well into a single curve (and this collapse occurs for any direction of $\bm{r}_{ij}$).
It means that the effective interaction is actually of the form of Eq.~(\ref{eq:J_scaling}) and the use of Eq.~(\ref{eq:C_J}) is indeed justified.
Usually, Kac's prescription is regarded as a purely mathematical operation 
to extract nontrivial thermodynamic properties of long-range interacting systems~\cite{Les_Houches2008}.
However, in the present model, such a mathematical operation is not necessary;
the scaling of Eq.~(\ref{eq:J_scaling}) naturally emerges in the effective interaction.
This is a remarkable characteristic of this model.

In Fig.~\ref{fig:elastic}, we compared some equilibrium quantities of the elastic model with those of the effective Ising model,
$H_{\rm eff}=-\sum_{ij}\hat{J}_{ij}\sigma_i\sigma_j-D\sum_i^N\sigma_i$.
In all the simulations presented in Fig.~\ref{fig:elastic}, 
we used the effective interaction $\hat{J}_{ij}$ estimated at $T=0.5$ in the canonical ensemble.
The numerical results of the effective Ising model are plotted by green triangles.
Although the effective interaction is estimated by the data at the single point in the canonical ensemble, 
the elastic model is indistinguishable from the effective Ising model for all the values of parameters and for all the statistical ensembles.

To summarize, we have argued that a certain short-range interacting system displays nonadditivity when the system is in a quasiequilibrium state,
although a genuine equilibrium state would be additive.
Within the time window~(\ref{eq:condition}), the system relaxes to a quasiequilibrium state, 
which is described by an equilibrium state of the elastic spin model~(\ref{eq:elastic_H}).
It has been found that the elastic spin model exhibits ensemble inequivalence.
In addition, it has been numerically shown that as far as equilibrium states of spin degrees of freedom are concerned,
the elastic spin model is indistinguishable from the effective Ising model with a long-range interaction.
It implies that statistical mechanics of long-range interacting systems is relevant 
for understanding quasiequilibrium states of some short-range interacting systems.
The limitation of the present study is that several time scales should be separated as Eq.~(\ref{eq:condition}), 
which requires the large potential depth $V_0$ compared to $k_{\rm B}T$; 
otherwise considering an equilibrium state of the Hamiltonian (\ref{eq:elastic_H}) would not be justified.
Finally, it is pointed out that the present study will give some insight into experimental attempts to realize a long-range interacting system 
and observe its peculiar properties in laboratory~\cite{ODell2000,Chalony2013}.
Such attempts are not satisfactory yet, but we hope that this work stimulates different experimental approaches toward it.

The author thanks Taro Nakada for fruitful discussion and Seiji Miyashita for continual discussion and careful reading of the manuscript.
He acknowledges the financial support provided by the Sumitomo Foundation.
The computation in this work has been done using the facilities of the Supercomputer Center, Institute for Solid State Physics, University of Tokyo.


\end{document}